\RequirePackage{fix-cm}
\documentclass[smallextended]{svjour3}       
\smartqed  
\usepackage{graphicx}
\begin{document}

\title{How to reduce the number of rating scale items without predictability loss?
}

\titlerunning{How to reduce the number of rating scale...}        

\author{W.W. Koczkodaj    \and
        T. Kakiashvili    \and
        A. Szyma\'nska    \and
        J. Montero-Marin  \and
        R. Araya          \and
        J. Garcia-Campayo \and
        K. Rutkowski      \and
        D. Strza\l{}ka 
}

\authorrunning{W.W. Koczkodaj et al.} 

\institute{W.W. Koczkodaj \at
           Computer Science, Laurentian University, Sudbury Campus, 935 Ramsey Lake Rd., Sudbury, ON P3E 2C6, Ontario, Canada \\
           \email{wkoczkodaj@cs.laurentian.ca}             \\
           \and
           T. Kakiashvili \at
           Sudbury Therapy, Ontario, Canada\\
           \and
           A. Szyma\'nska \at
           UKSW University, Dewajtis 5, 01-815 Warszawa, Poland \\
           \and
           J. Montero-Marin \at
           Faculty of Health Sciences and Sports, University of Zaragoza, Spain\\
           \email{jmonteromarin@hotmail.com}\\
           \and
           R. Araya \at
           London School of Hygiene and Tropical Medicine, Centre for Global Mental Health, United Kingdom\\
           \and
           J. Garcia-Campayo \at
           Miguel Servet Hospital, University of Zaragoza, Spain\\
           \and
           K. Rutkowski \at
           Jagiellonian University, Go{\l}\c{e}bia 24, 31-007 Krak\'ow, Poland\\
           \and
           D. Strza{\l}ka \at
           Faculty of Electrical and Computer Engineering, Rzesz\'ow University of Technology, Al. Powsta\'nc\'ow Warszawy 12, 35-959 Rzesz\'ow, Poland\\
           \email{strzalka@prz.edu.pl}
}

\date{Received: date / Accepted: date}

\maketitle

\begin{abstract}

Rating scales are used to elicit data about qualitative entities (e.g., research collaboration). This study presents an innovative method for reducing the number of rating scale items without the predictability loss. The ``area under the receiver operator curve method'' (AUC ROC) is used. The presented method has reduced the number of rating scale items (variables) to 28.57\% (from 21 to 6) making over 70\% of collected data unnecessary. 

Results have been verified by two methods of analysis: Graded Response Model (GRM) and Confirmatory Factor Analysis (CFA). GRM revealed that the new method differentiates observations of high and middle scores. CFA proved that the reliability of the rating scale has not deteriorated by the scale item reduction. Both statistical analysis evidenced usefulness of the AUC ROC reduction method.

\end{abstract}

\section{Introduction}
\label{intro}

Rating scales (also called assessment scale) are used to elicit data about quantitative entities (e.g., research collaboration as in \cite{BMD2009}). Often, predictability of rating scales (also called ``assessment scales'') could be improved. 
Rating scales often use values: ``1 to 10'' and some rating scales may have over 100 items (questions) to rate.  
Other popular terms for rating scales are: \textit{survey} and \textit{questionnaire} although a questionnaire is a method of data collection while survey may not necessarily be conducted by questionnaires. Some surveys may be conducted by interviews or by analyzing web pages. Rating itself is very popular on the Internet for ``Customer Reviews'' where often uses five stars (e.g., by Amazon.com) instead of ordinal numbers. One may regard such rating as a one item rating scale. Surveys are used in \cite{CW2016} on Fig.1 (with the caption: ``Sketch of data integration in use for different purposes with interference points for standardisation'') as one of the main sources of data. 

A survey, based on the questionnaire, answered by 1,704 researchers from 86 different countries, was conducted by the Scientometrics study \cite{BZ2012} on the impact factor, which is regarded as a controversial metric. Rating scales were also used in \cite{P2007} and \cite{KKL2014}. In \cite{KKW2011,GKK2013}, a different type of the rating scale improvement was used (based on pairwise comparisons). The evidence of improving accuracy by pairwise comparisons is in \cite{Kocz96} and \cite{Kocz98}.

According to \cite{MR2012}: 
\begin{quotation}
	... the differentiation of sciences can be
	explained in a large part by the diffusion of generic instruments created by research-technologists
	moving in interstitial arenas between higher education, industry, statistics
	institutes or the military. We have applied this analysis to research on depression by
	making the hypothesis that psychiatric rating scales could have played a similar role in the development of this scientific field.
\end{quotation}

The absence of a well-established unit (e.g., one kilogram or meter) for measuring the science compels us to use rating scales. They have great application to scientometrics for measuring and analyzing performance based on subjective assessments. Even granting academic degrees is based on rating scales (in this case, several exams which are often given to students by questionnaires). Evidently, we regard this rating scale as accurate otherwise our academic degrees may not have much value.

The importance of subjectivity processing was driven by the idea of \textit{bounded rationality}, proposed by Herbert A. Simon (the Nobel Prize winner), as an alternative basis for the mathematical modelling of decision making. 

\section{The data model}
\label{sec:1}

Data collected by a rating scale with fixed number of items (questions) are stored in a table with one decision (in our case, binary) variable. The parametrized classifier is usually created by total score of all items. Outcome of such rating scales is usually compared to external validation provided by assessing professionals (e.g., grant application committees).

Our approach not only reduces the number of items but also sequences them according to the contribution to predictability. It is based on the Receiver Operator Characteristic (ROC) which gives individual scores for all examined items.

\subsection{Predictability measures}
\label{sec:2.1}

The term ``receiver operating characteristic'' (ROC), or ``ROC curve'' was coined for a graphical plot illustrating the performance of radar operators (hence ``operating''). A binary classifier represented absence or presence of an enemy aircraft and was used to plot the fraction of true positives out of the total actual positives (TPR = true positive rate) vs. the fraction of false positives out of the total actual negatives (FPR = false positive rate).
Positive instances (P) and negative instances (N) for some condition are computed and stored as four outcomes a 2 \time 2 contingency table or confusion matrix, as follows:

\begin{table}[htbp]
	\centering
	\caption{The confusion matrix}
	\begin{tabular}{cc}
		\hline
		\textbf{True Positives} & \textbf{False Positives} \\
		\hline
		\textbf{False Negative} & \textbf{True Negative} \\
		\hline
	\end{tabular}%
	\label{tab:c-m}%
\end{table}%

In assessment and evaluation research, the ROC curve is 
a representation of a ``separator'' (or decision) variable. The decision variable is usually: ``has a property'' or ``does not have a property'' or has some condition to meet (pass/fail).
The frequencies of positive and negative cases of the diagnostic test vary for the ``cut-off'' value for the positivity. By changing the ``cut-off'' value from 0 (all negatives) to a maximum value (all positives), we obtain 
the ROC by plotting TPR (true positive rate also called sensitivity) versus FPR (false positive also called specificity) across varying cut-offs, which generate a curve in the unit square called an ROC curve. 

According to \cite{TF2006}, the area under the curve (the AUC or AUROC) is equal to the probability that a classifier will rank a randomly chosen positive instance higher than a randomly chosen negative one (assuming the 'positive' rank higher than 'negative'). 

AUC is closely related to the \textit{Mann-Whitney U test} which tests whether positives are ranked higher than negatives. It is also equivalent to the Wilcoxon test of ranks. The AUC is related to the Gini coefficient given by the formula 
\begin{equation}
\label{eq1}
G_1 = 2*AUC - 1,
\end{equation}

where:
$$G_1 = 1 - \sum_{k=1}^n (X_{k} - X_{k-1}) (Y_k + Y_{k-1})$$

In this way, it is possible to compute the AUC using an average of a number of trapezoidal approximations.
Practically, all advanced statistics can be questioned and they often gain recognition after their intensive use. The number of publications with ROC listed by PubMed.com has exploded in the last decade and reached 3,588 in 2013. An excellent tutorial-type introduction to ROC is in \cite{TF2006}. It was introduced during the World War II for evaluation of performance the radar operators. 
Its first use in health-related sciences, according to Medline search, is traced to \cite{CJ1967}.

\begin{table}[htbp]
	\centering
	\caption{AUC of individual variables in the original data}
	\begin{tabular}{rrrrrr}
		\hline
		Var   & AUC   & Var   & AUC   & Var   & AUC \\
		\hline
		21    & 0.587468 & 12    & 0.636791 & 17    & 0.674283 \\
		11    & 0.597342 & 13    & 0.648917 & 15    & 0.692064 \\
		16    & 0.605937 & 4     & 0.651187 & 10    & 0.697225 \\
		6     & 0.610004 & 3     & 0.655666 & 9     & 0.700461 \\
		18    & 0.610028 & 5     & 0.658478 & 7     & 0.701489 \\
		19    & 0.629285 & 20    & 0.666999 & 14    & 0.707401 \\
		2     & 0.631205 & 8     & 0.667983 & 1     & 0.725009 \\
		\hline
	\end{tabular}%
	\label{tab:AUC}%
\end{table}%

\subsection{Validation of the predictability improvement}
\label{sec:2.2}

Supervised learning is the process of inferring a decision (of classification) from labeled training data. However, the supervised learning may also employ other techniques, including statistical methods that summarize and explain key features of the data. 
For the unsupervised learning, clustering is the most popular method for analyzing data. The k-means clustering optimizes well for the given number of classes. In our case, we have two classes: 0 for ``negative'' and 1 for ``positive'' outcome of diagnosis for depression. 

The area under the receiver operating characteristic curve (AUC) reflects the relationship between sensitivity and specificity for a given scale. An ideal scale has an AUC score equal to 1 but it is not realistic in clinical practice. Cutoff values for positive and negative tests can influence specificity and sensitivity, but they do not affect AUC. The AUC is widely recognized as the performance measure of a diagnostic test's discriminatory power (see \cite{LBZO2005,ZOM2007}). In our case, the input data have AUC of 81.17\%.

The following System R code was used to compute the AUC for all 21 individual items:

\begin{minipage}[t]{4.75in} 
	
	
	library(caTools) \\
	\# read data from a csv file \\
	mydata = read.csv("C:
	\textbackslash\textbackslash BDI571.csv") 
	
	y = mydata[,1] \\
	result<-matrix(nrow=22,ncol=2); \\
	ind=2; \\
	
	for (i in 2:22) \\ \{ \\
	\hspace*{18pt} result[ind,]=colAUC(cbind(mydata[,1], \\
	\hspace*{36pt} mydata[,i]),y, plotROC=FALSE, alg="ROC") \\
	\hspace*{18pt} ind = ind+1 \\
	\} \\

	System R code \\
\end{minipage}

\begin{figure}
	\centering
	\includegraphics[width=1\linewidth]{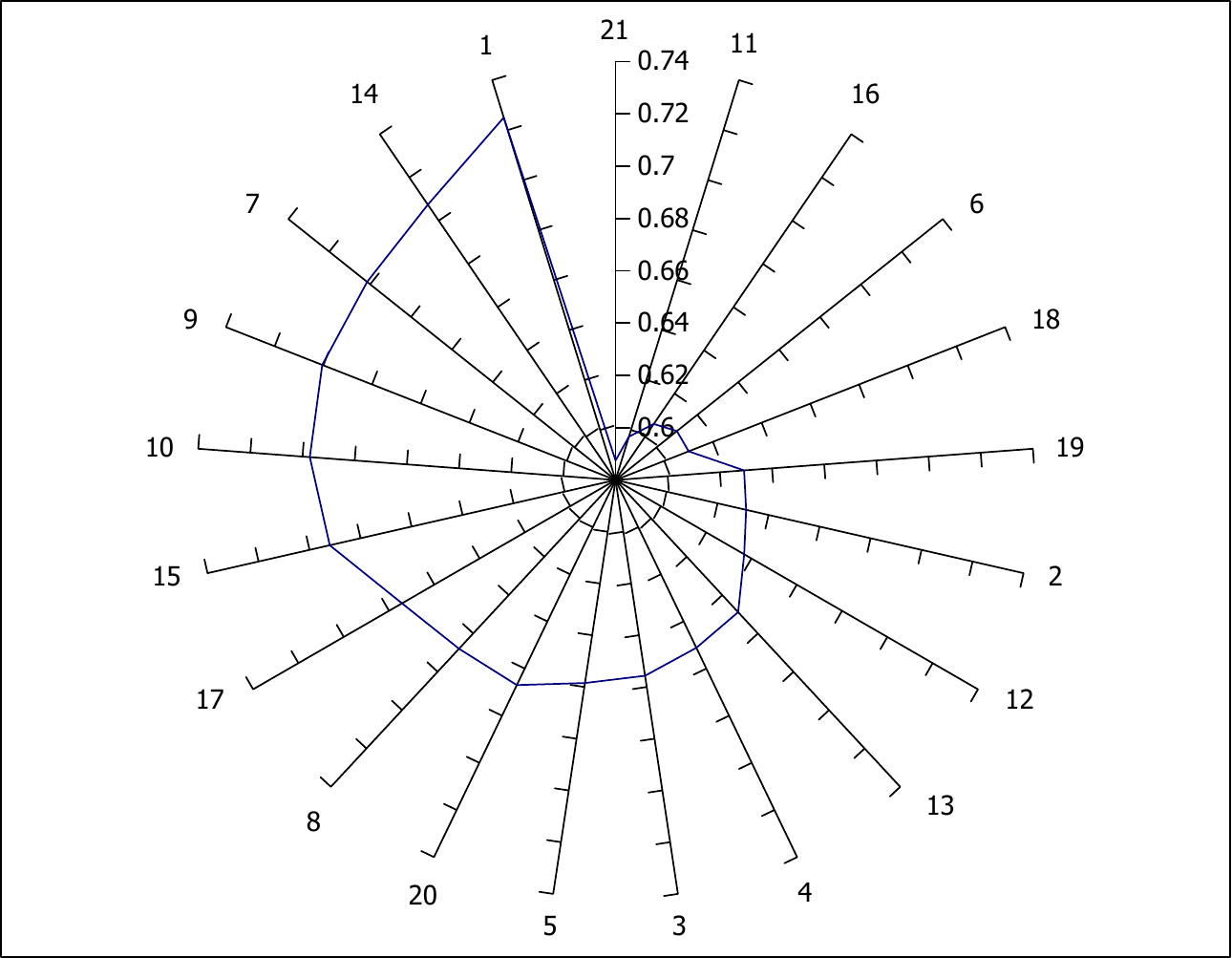}
	\caption{AUC for the running total of all variables}
	\label{fig:AUC-graph1}
\end{figure}

When AUC values are computed for all individual variables, we arrange them in an ascending order. These variables are present in Table~\ref{tab:AUC2} in bold. Values in the row below running total up to the current variable. Evidently, the first value 0.725 is the same as in 
Table~\ref{tab:AUC} since the running total is the single variable 1. However, the third value in the second row (0.795) is not for variable 7 but the total of variables 1, 14, and 7. In particular, the last value (0.812) in Tab.~\ref{tab:AUC2} is for the total of all variables. Frankly, these numbers are very close to each other but their line plot 
\ref{fig:AUC-graph1} demonstrates its usefulness. The curve peek is for variable \#6 which is 15. There is a slight decline until variable 16. 

\begin{table}[htbp]
	\centering
	\caption{AUC of running variable totals}
	\begin{tabular}{rrrrrrr}
		
		\hline
		\textbf{1} & \textbf{14} &\textbf{7} &\textbf{ 9} & \textbf{10}  & \textbf{15}    & \textbf{17} \\
		0.725 & 0.777 & 0.795 & 0.810 & 0.813 & 0.822 & 0.821 \\
		\textbf{8} & \textbf{20} & \textbf{5} & \textbf{3} & \textbf{4}     & \textbf{13}    & \textbf{12} \\
		0.819 & 0.820 & 0.821 & 0.821 & 0.821 & 0.821 & 0.820 \\
		\textbf{2} & \textbf{19} & \textbf{18} & \textbf{6} & \textbf{16}    & \textbf{11}    & \textbf{21} \\
		0.819 & 0.818 & 0.816 & 0.814 & 0.812 & 0.811 & 0.812 \\
		\hline
	\end{tabular}%
	\label{tab:AUC2}%
\end{table}%

\section{Relating the results to Graded Response Model}
\label{sec:3}

Let us examine how our results can be related to the Graded Response Model (GRM). GRM is equivalent of Item Response Theory, well addressed by a Wikipedia article,
but used for ordinary, not binary, data.
GRM is usually conducted to establish the usefulness of test items \cite{ADK1992}. 

GRM is used in psychometric scales to determine the level of three characteristics of each item, namely: a) item's difficulty, b) item's discriminant power, and c) item's guessing factor.

\textbf{Item's difficulty} describes how difficult or easy it is for individuals to answer on the item. High positive value means that the item is very difficult, high negative value means that the item is very easy. 

\textbf{  Item's discriminant power} describes ability for a specific item to distinguish among upper and lower ability individuals' on a test.

\textbf{Item's guessing factor }describes probability that individual with low feature (low depression) achieved high scores in this item.

The aim of our analysis was to establish whether or not the GRM indicates the same items as the proposed method based on AUC. Two GRM models were build for the given rating scales:

\textbf{constrained} (that assumes equal discrimination parameters across items),\\

\textbf{unconstrained} (that assumes unequal discrimination parameters across items).\\

\begin{figure}[h]
	\centering
	\includegraphics[width=0.7\linewidth]{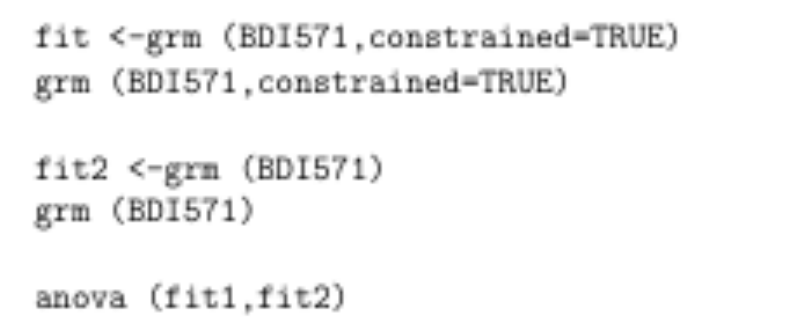}
	\caption{System R code for GRM models.}
	\label{fig:printscreen2}
\end{figure}

System R \textit{ltm} package \cite{R2006} was used in our analysis. Fig.~\ref{fig:printscreen2} illustrates system R code for GRM models.\

In order to check whether or not the unconstrained GRM provides a better fit than the constrained GRM, a likelihood ratio test was used. It revealed that unconstrained GRM is preferable (fit2 in Tab.~\ref{tab:LR}). The results of the Likelihood Ratio are presented in Tab.~\ref{tab:LR}. \\

\begin{table}[h]
	\centering
	\caption{Likelihood Ratio for the full GRM model}
	\label{tab:LR}
	\begin{tabular}{|c|c|c|c|c|c|c|} 
		\hline & AIC  &    BIC &  log.Lik  &  LRT & df & p-value \\
		\hline fit1 & 25494.12 & 25772.35 & -12683.06 &  & &\\               
		fit2 & 25367.63 & 25732.81 & -12599.81 & 166.49 & 20  & p $<$ 0.001\\
		\hline
	\end{tabular}
\end{table}

Tab.~\ref{tab:GRM} shows the unconstrained GRM model results with the item discrimination power. It provides information on discrimination power of each item. \\

\begin{table}[h]
	\centering
	\caption{Unconstrained GRM model results for the full rating scale and the item discrimination power}
	\label{tab:GRM}
	\begin{tabular}{|c|c|c|c|c|}
		\hline & Extrmt1 & Extrmt2 & Extrmt3 & Dscrmn  \\
		\hline V1   &  \textbf{0.178} &  \textbf{ 1.099}  &  \textbf{2.542} & \textbf{ 1.799}  \\
		V2  &   0.214  &  1.536   & 2.485 &  1.315  \\
		V3   & -0.094  &  1.306   & 3.077 &  1.528  \\
		V4   & -0.447  &  1.673  &  3.511 & 1.268    \\
		V5   & -0.709  &  1.903  &  2.717 &  1.440    \\
		V6   & -0.073  & 1.679  &  2.194 &  0.860   \\
		V7   & \textbf{-0.410 } &  \textbf{0.970}  & \textbf{ 2.170}  &\textbf{ 1.459}  \\
		V8   & -0.641  &  0.876  &  2.157 &  1.461  \\
		V9   &  \textbf{0.092}  & \textbf{ 1.895}  &  \textbf{2.471} & \textbf{ 1.405}  \\
		V10  & \textbf{-0.183 } & \textbf{ 0.834  }&  \textbf{1.665} &  \textbf{1.023}   \\
		V11  & -0.881  &  2.187  &  3.280 &  0.767   \\
		V12  & -0.242  &  1.282  &  2.221 &  1.271  \\
		V13  & -0.351  &  1.631  &  2.660  & 1.054  \\
		V14  & \textbf{-0.038}  &  \textbf{0.918}  & \textbf{ 2.353} & \textbf{ 1.951}   \\
		V15  & \textbf{-0.627 } &  \textbf{1.046 } & \textbf{ 2.248 } &\textbf{ 1.593}  \\
		V16  & -2.364  &  0.634  &  2.388  & 0.764  \\
		V17  & -0.482  &  1.287  &  2.366  & 1.296  \\
		V18  & -1.685  &  0.847  &  2.024  & 0.902   \\
		V19  & -1.623  &  0.366  &  2.648  & 1.078  \\
		V20  & -0.575  &  1.227  &  2.066 &  1.643 \\
		V21  &  1.271  &  2.240  &  3.531  & 0.870  \\
		\hline 
	\end{tabular} 
\end{table}

Items selected by AUC ROC are shown in Tab.~\ref{tab:GRM} as bold. Evidently, they have the large discrimination power (seen in the last column). All selected items discriminate between responses above the mean value (so on their basis we can discriminate between respondents with severe and moderate level of depression). Discrimination power is 
a characteristic of items in the scale. It is a measurement method which aim is to assess how respondents differ in their answers on rating scale items. The larger is 
the discrimination power of the item, the better, more useful is item in the scale \cite{Anastasi}. Items computed by the proposed (AUC ROC) method have a good discrimination as it can be seen in the Table 4 (for example, number 1.799 means that item $V1$ has a good discrimination power).

All items of the given rating scale give 56.21\% of total information for the latent trait
and the latent variable (adolescent depression in school in our case). Test Information Curve (see Tab.~\ref{tab:InfoCurve}) shows that six items provides 19.62\%  of the total information for latent trait. The higher is items' discrimination, the more information or precision the scale provides.    

GRM model computes different items than our proposed method. AUC ROC is based on the count of true and false positive rate while GRM model is based on the maximum likelihood estimate. The proposed method has a bigger diagnostic power. Diagnostic power is 
the ability of the test to detect all subjects, which have 
been measured by the test characteristics (in our case, for depression). A test with
the maximum diagnostic power would detect all subjects (suffering from depression). Unfortunately, the most selections of rating scale items do not compare solutions with the diagnostic criterion. That is why the proposed method is so useful for the selection of items in different measurement tools (examination, tests, socio-metrical scales, psychometrical scales, and many others).

We used GRM model here to show that even such powerful method like GRM (used in psychometrics to indicate which items can discriminate subjects), does not provide an answer to a question about diagnostic accuracy of items. According to GRM items, V2 and V3 (Table 5) have a considerable discriminant power, but the proposed method shows which items better discriminate between subjects on the basis of diagnostic criteria.

\begin{table}[t]
	\centering
	\caption{Test information curve.}
	\label{tab:InfoCurve}
	\begin{tabular}{|c|}
		\hline Total information = 56.21\\ 
		Information in (-4, 4) = 52 (92.51\%)\\
		Based on all the items\\
		\hline  Total Information = 19.62\\
		Information in (-4, 4) = 18.97 (96.65\%)\\
		Based on items 1, 7, 9, 10, 14, 15\\
		\hline 
	\end{tabular} 
\end{table}

\section{Reduced scale psychometric properties}
\label{sec:4}

Confirmatory Factor Analysis (CFA) \cite{Hair2006} \cite{BSMG2008} was used to verify the structure of our results. CFA is a factor analysis which purpose is to verify the structural validity
whether items belong to scales and what are their factor loading. Factor loading measures the relations between observed variable (item) and latent feature (scale).
The higher the factor loading, the stronger the relation, and the item has greater importance in the scale. More specifically, CFA was used to determine whether: 
\begin{itemize}
	\item 
	items indicated by AUC form a coherent scale that exhibits good reliability,
	\item 
	the reliability of the rating scale has not deteriorated by the scale item reduction.
\end{itemize}

Two CFA models were built. The first CFA model has all items and the second CFA model has a reduced number of items. Since items of the scale have categorical format, the robust estimator WLSMV (weighted least squares means and variance, see \cite{BHP2006}) was used as it is designed for categorical scales. The robust estimator resists the lack of normal distributions. The analysis was conducted in ``lavaan'' package of R program.\

\begin{figure}[h]
	\centering
	\includegraphics[width=1.0\linewidth]{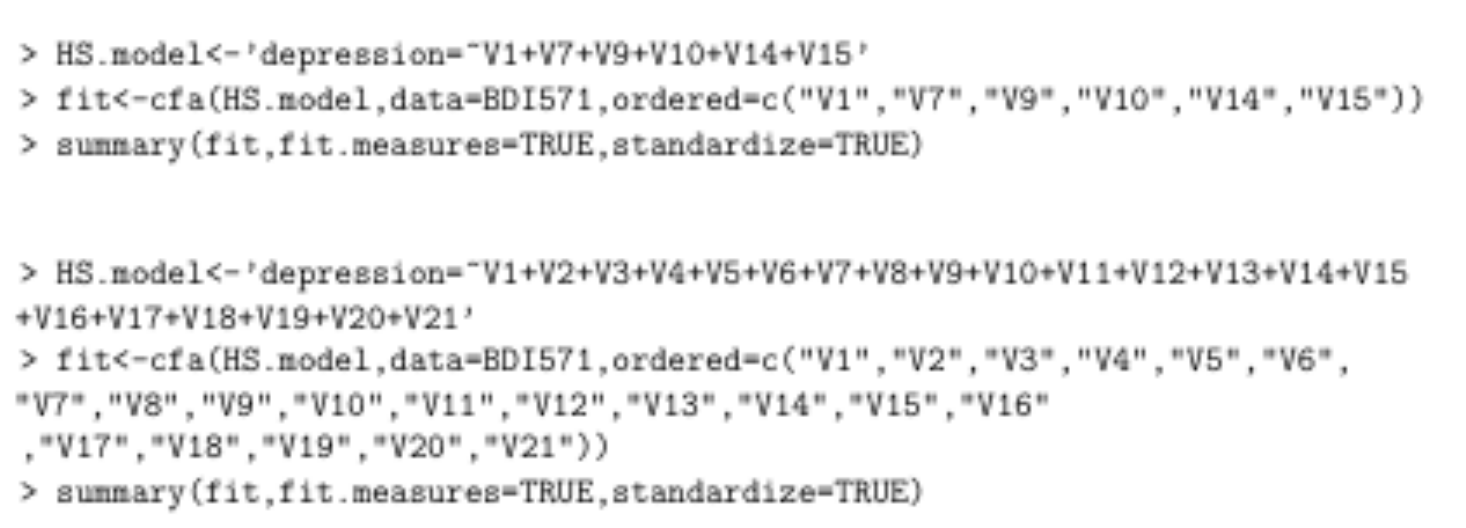}
	\caption{System R code for CFA}
	\label{fig:printscreen}
\end{figure}

\begin{figure}[h]
	\centering
	\includegraphics[width=0.9\linewidth]{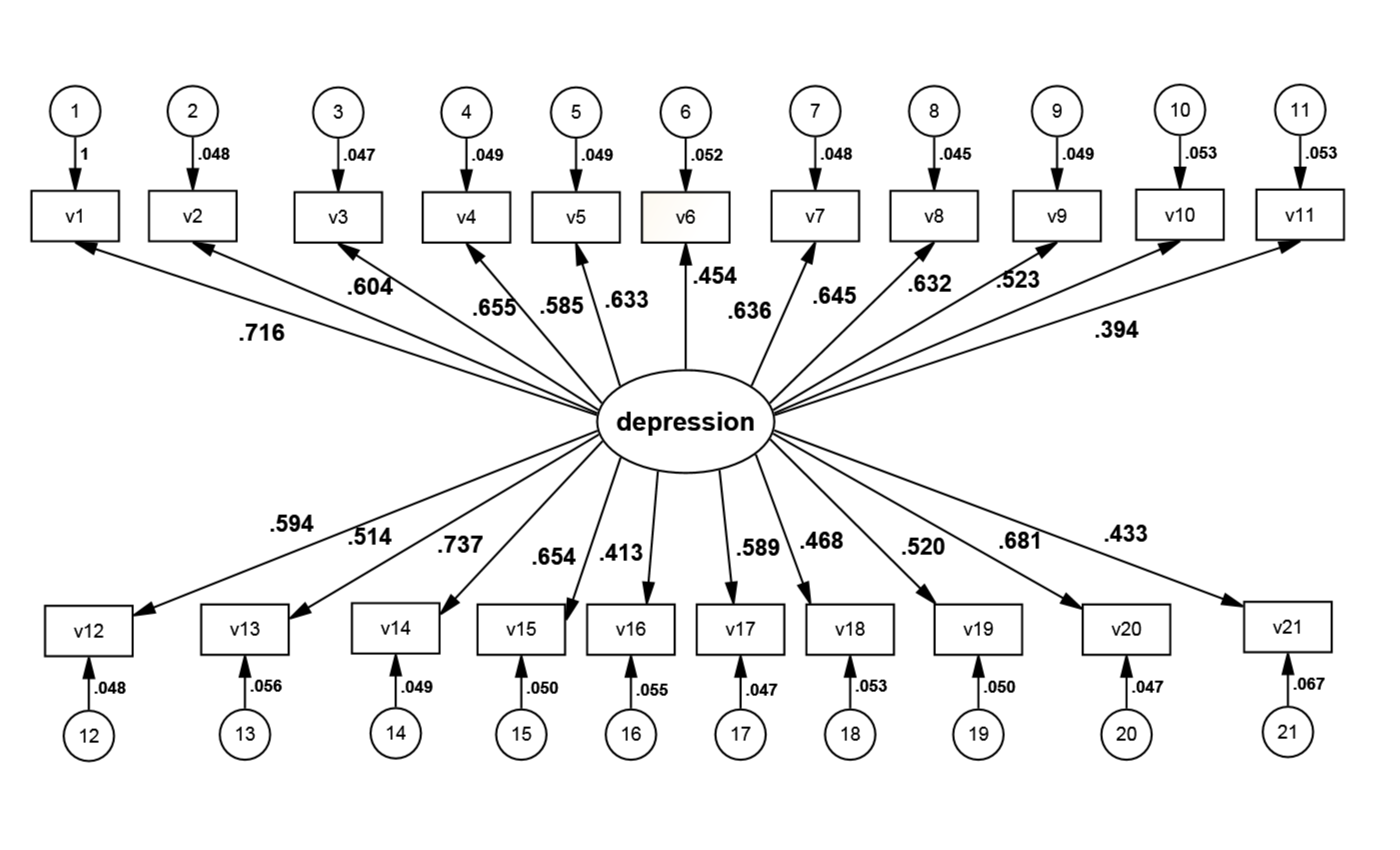}
	\caption{CFA  model for the rating scale with all items presented in AMOS graphics}
	\label{fig:deplong}
\end{figure}

\begin{table}[h]
	\centering
	\caption{Parameter estimates of the full rating scale}
	\label{tab:parEst}
	\begin{tabular}{|c|c|c|c|}
		\hline parameters & standardized & non-standardized & standardized error \\ 
		\hline ${\lambda }_{V1}$ & \textbf{0.716} & 1.000 &  \\ 
		${\lambda }_{V2}$ & 0.604 & 0.844 & 0.048   \\
		${\lambda }_{V3}$ & 0.655 & 0.914 & 0.047\\
		${\lambda }_{V4}$ & 0.585 & 0.818 & 0.049 \\
		${\lambda }_{V5}$ & 0.633 & 0.884 & 0.049\\
		${\lambda }_{V6}$ & 0.454 & 0.634 & 0.052\\
		${\lambda }_{V7}$ & \textbf{0.636} & 0.889 & 0.048\\
		${\lambda }_{V8}$ & 0.645 & 0.901 & 0.045\\
		${\lambda }_{V9}$ & \textbf{0.632} & 0.883 & 0.049\\
		${\lambda }_{V10}$ & \textbf{0.523} & 0.731 & 0.053\\
		${\lambda }_{V11}$ & 0.394 & 0.550 & 0.053\\
		${\lambda }_{V12}$ & 0.594 & 0.830 & 0.048\\
		${\lambda }_{V13}$ & 0.514 & 0.718 & 0.056\\
		${\lambda }_{V14}$ & \textbf{0.737} & 1.029 & 0.049\\
		${\lambda }_{V15}$ & \textbf{0.654} & 0.913 & 0.050\\
		${\lambda }_{V16}$ & 0.413 & 0.578 & 0.055\\
		${\lambda }_{V17}$ & 0.589 & 0.822 & 0.047\\
		${\lambda }_{V18}$ & 0.468 & 0.653 & 0.053\\
		${\lambda }_{V19}$ & 0.520 & 0.726 & 0.050\\
		${\lambda }_{V20}$ & 0.681 & 0.952 & 0.047\\
		${\lambda }_{V21}$ & 0.433 & 0.605 & 0.067\\
		\hline 
	\end{tabular}  
\end{table}

The model for the full rating scale is presented by Fig.~\ref{fig:deplong}.
Table \ref{tab:parEst} presents parameter estimates of the full rating scale. Loads of those items, which have been identified by the presented method as having the greatest predictive power, is in bold in Tab.~\ref{tab:parEst}.
A model with a reduced number of items is in Fig.~\ref{fig:deplong}. Tab.~\ref{tab:EstimRedSc} presents parameter estimates for the reduced scale model.\\

\begin{figure}[h]
	\centering
	\includegraphics[width=1\linewidth]{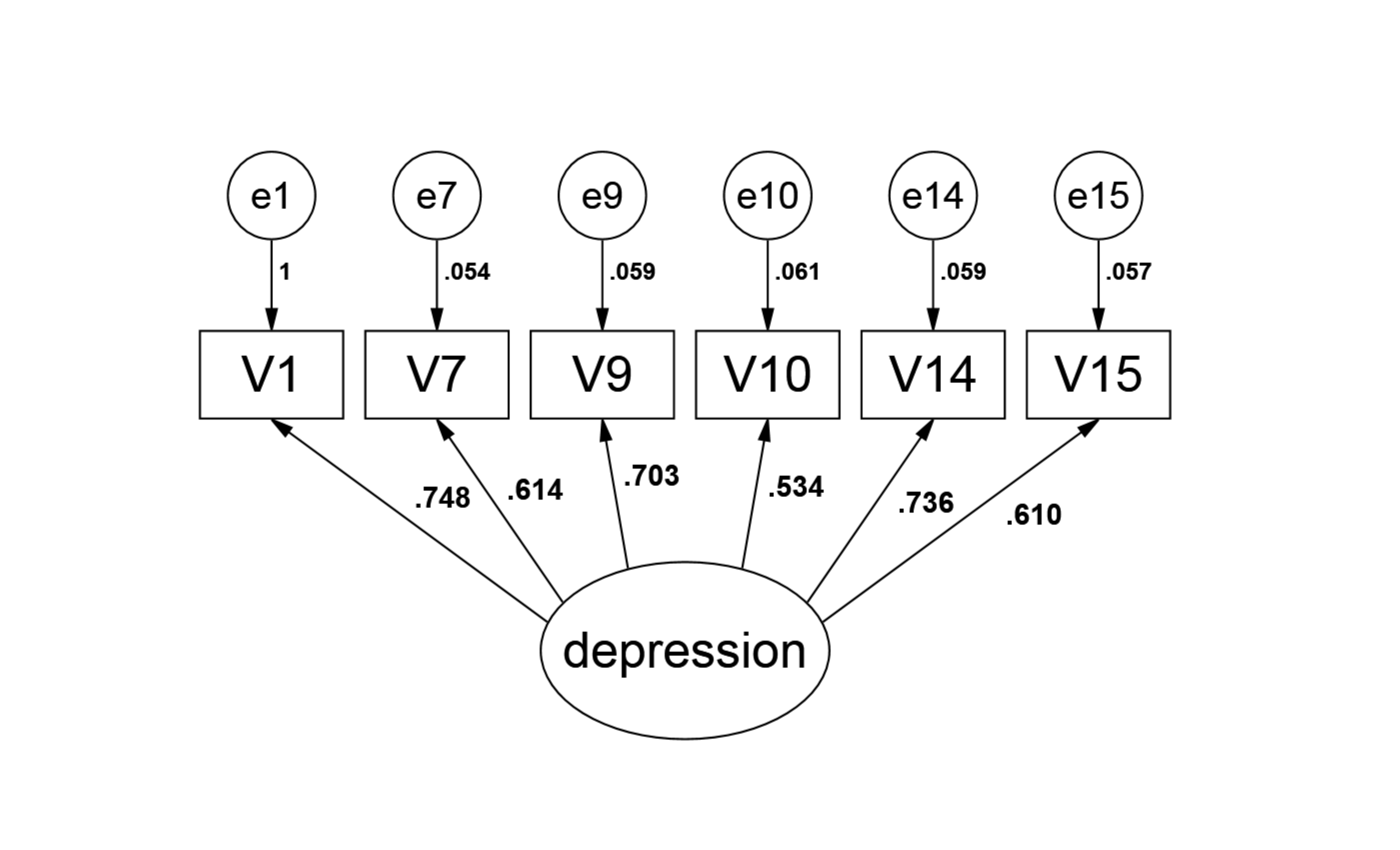}
	\caption{CFA Model with a reduced number of items presented in AMOS graphics}
	\label{fig:depshort}
\end{figure}

\begin{table}[h]
	\centering
	\caption{Parameter estimates for the reduced scale}
	\label{tab:EstimRedSc}
	\begin{tabular}{|c|c|c|c|}
		\hline parameters & standardized & non-standardized & standardized error \\ 
		\hline ${\lambda }_{V1}$ & 0.748 & 1.000 &  \\ 
		${\lambda }_{V7}$ & 0.614 & 0.821 &   0.054 \\
		${\lambda }_{V9}$ & 0.703 & 0.940 & 0.059\\
		${\lambda }_{V10}$ & 0.534 & 0.714 & 0.061\\
		${\lambda }_{V14}$ & 0.736 & 0.984 & 0.059\\
		${\lambda }_{V15}$ & 0.816 & 0.816 & 0.057\\
		
		\hline 
	\end{tabular}  
\end{table}

For the purpose of checking whether the models have 
a good fit, we used two fit indices: CFI (cross validation index) and RMSEA (root mean square error of approximation). According to \cite{BSMG2008,SSV2009}, both CFA models have a good fit to the data as illustrated by Tab.~\ref{tab:FitStat}. Values of CFI  statistics for both models exceeded the required level of 0.9.  For both models, the values of RMSEA  statistics (lower than 0.08) indicates the good fitness of the proposed new scale structure for the given data.\\

\begin{table}[h]
	\centering
	\caption{Results of fit statistics for two rating scale models}
	\label{tab:FitStat}
	\begin{tabular}{|c|c|}
		\hline 
		Statistics for the full and reduced rating scale models & \\ 
		\hline Chi2=437.899 & Chi2=30.883 \\ 
		\hline df=189 & df=9 \\ 
		\hline CFI=.950 & CFI=.983 \\ 
		\hline RMSEA=.048 & RMSEA=.065 \\ 
		\hline 
	\end{tabular}
\end{table}

For both CFA models, \textit{construct reliability} (CR) and \textit{variance extracted} (VE) 
were computed. $CR$ was computed by the formula (given in \cite{Hair2006}):\\

\begin{equation}
\label{eq2}
 CR = \frac{{\left(\sum _{i=1}^{n}{\lambda }_{i}\right)}^{2}}{{\left(\sum _{i=1}^{n}{\lambda }_{i}\right)}^{2}+\left(\sum _{\stackrel{.}{c}=1}^{n}{\delta }_{i}\right)}
\end{equation}

\noindent where:\\
\quad \textit{i} is a total number of items,\\
\quad $\lambda $ is a factor loading,\\
\quad $\delta $ is an error variance, which is the amount of variability unexplained by the items in scale.\\

\medskip
\noindent The formula for computing \textit{variance extracted} (VE) is based on \cite{Hair2006}:\\
\begin{equation}
\label{eq3}
VE = \frac{\sum _{i=1}^{n}{\lambda }_{i}^{2}}{n}
\end{equation}

\noindent where:\\
\quad \textit{i} is the number of items,\\
\quad $\lambda $ is a factor loading,\\
\quad \textit{n} is a number of rating scale items.\\
\bigskip

\begin{table}[h]
	\centering
	\caption{Results of CR and VE of two models}
	\label{tab:tabCRVR}
	\begin{tabular}{|c|c|c|}
		\hline Rating scale  & CR & VE \\ 
		\hline Full rating scale & 0.929 & 0.394 \\
		Reduced scale & 0.822 & 0.483 \\ 
		\hline 
	\end{tabular} 
\end{table}

The results revealed that the reliability of the reduced model CR = .822 and is lower than the reliability of the full model of 0.1 (CR=.929). Therefore, it can be concluded that the reliability of the scale is above the acceptability level. Removing 15 items has not impaired its reliability as Tab.~\ref{tab:tabCRVR} demonstrates it.

For the reduced model, VE=.438 while for the given model, VE=.394. Evidently, the new model has VE closer to criterion of .500. The reduced rating scale model has a better VE than the full rating scale model. It means that the reduced rating scale model explains the diversity of the results better than the full rating scale model (see Tab.~\ref{tab:tabCRVR}).

On the basis of factor loadings ($\lambda $), we are unable to determine which items have the most predictive power. Items V3 or V20 have one of the top factor loadings in the full rating scale, but they do not still have the most predictive power. Therefore, it is impossible to indicate the ordinal number of the rating scale item according to the factor analyses, but it is possible by the proposed method and GRM. However, GRM cannot compare its solution with a diagnostic criterion while the proposed method can.

\section{Discussion}
\label{sec:5}

The Beck Depression Inventory (BDI) was selected for our study since it is one of the best known and most widely used self-rating scales to assess the presence and severity of depressive symptoms \cite{BB1996,DSM,ICD}. Our data were collected in high schools \cite{AMF2011,AMB2013}.  
However, it needs to be stressed that our method is applicable to practically all rating scales.

In summary, both models fit the data well. Both of them have a good reliability and a relatively good variance. Reducing the number of items did not burden psychometric properties, but simplified the whole structure (as indicated by the smaller number of degrees of freedom). According to the Occam's Razor law, the simpler models, the better.
Although it was not the main objective of this study, it is worth to notice that the six rating scale items have a better predictive power in our study than other 21 items.
We have also demonstrated that our results have the domain (semantic) meaning.

\section{Conclusions}
\label{sec:6}

The presented method has reduced the number of the rating scale items (variables) to 28.57\% (from 21 to 6) making over 70\% of collected data unnecessary. It is not only an essential budgetary saving, as data collection is usually expensive, but it often contributes to the data collection error reduction. The more data are collected, the more errors are expected to occur. When we use the proposed AUC ROC reduction method, the predictability has increased by approximately 0.5\%. It may seem insignificant but for a large population, it is not so. In fact, \cite{WHO} states that: ``Taken together, mental, neurological and substance use disorders exact a high toll, accounting for 13\% of the total global burden.'' 

The proposed use of AUC for reducing the number of rating scale items is innovative and applicable to practically all rating scales. System R code is posted on the Internet for the general use. A package for System R is under development. Certainly, more validation cases would be helpful and the assistance will be provided to anyone who wishes to try this method using his/her data.

\section*{Supporting Information}

The source code will be deposited at SourceForge.net hosting provider (see \cite{sf}). According to \cite{sf}, SourceForge ``creates powerful software in over 400,000 open source projects ans hosts over 3.7 million registered users''. It connects well over 40 million customers with  more than 4,800,000 downloads a day.

\section*{Acknowledgments}
The first author was supported (in part) by the Euro Research grant ``Human Capital''. 
Authors would like to thank Prof. E. Aranowska, specialized in psychometry, for reading the first draft.
Authors would like to thank Amanda Dion-Groleau (Laurentian University, Psychology) and Grant O. Duncan, Team Lead, Business Intelligence and Software Integration, Health Sciences North, Sudbury, Ontario for their help with proofreading this text.


\begin{thebibliography}{}

\bibitem{Anastasi}
Anastasi, A., Urbina, S.  Testy psychologiczne. Warszawa: Pracownia Test\'ow Psychologicznych Polskiego Towarzystwa Psychologicznego, (1999).	

\bibitem{AMF2011}
Araya, R., Montgomery, A. A., Fritsch, R., Gunnell, D., Stallard, P., Noble, S., Martinez, V., Barroilhet, S., Vohringer, P., Guajardo, V., 
School-based intervention to improve the mental health of low-income, secondary school
students in Santiago, Chile (YPSA): study protocol for a randomized controlled trial. Trials, vol. 12, p. 49, (2011).

\bibitem{AMB2013}
Araya, R., Montero-Marin, J., Barroilhet, S., Fritsch, R., Montgomery, A.,
Detecting depression among adolescents in Santiago, Chile: sex differences. BMC Psychiatry, vol. 13, p. 269, (2013).

\bibitem{ADK1992}
Ayala, R. J. D., Dodd, B. G., Koch, W. R.,  
A Comparison of the Partial Credit and Graded Response Model in Computerized Adaptive Testing. 
Applied Measurement in Education, vol. 5(1), pp. 17-–34, (1992).

\bibitem{BSMG2008}
Bartholomew, D. J., Steele, F., Moustaki, I., Galbraith, J. I.,  Analysis of multivariate social science data. Boca Raton, FL: Chapman \& Hall/CRC Press, (2008).

\bibitem{BHP2006}
Beauducel, A., Herzberg, P. Y.,
On the Performance of Maximum Likelihood versus Means and Variance Adjusted Weighted Least Squares Estimation in CFA Structural Equation. Modeling: A Multidisciplinary Journal, vol. 13(2), pp. 186--203, (2006).

\bibitem{BB1996}
Beck, A. T., Steer, R. A., Brown, G. K., BDI-II - Beck Depression Inventory Manual. Volume Second. San Antonio: The Psychological Corporation, (1996).

\bibitem{BZ2012}
Buela-Casal, G., Zych, I., 
What do the scientists think about the impact factor? 
Scientometrics, vol. 92(2), pp. 281--292, (2012).

\bibitem{BMD2009}
Bornmann, L., Mutz, R., Daniel, H. D., 
The influence of the applicants' gender on the modelling of a peer review process by using latent Markov models. Scientometrics, vol. 81(2), pp. 407--411, (2009).

\bibitem{CW2016}
Cinzia, D., Wolfgang, G., Grand challenges in data integration—state of the art
and future perspectives: an introduction. Scientometrics, vol. 108, pp. 391-–400, (2016).

\bibitem{CJ1967}
Carterette, E. C., Jones, M. H., 
Visual and auditory information processing in children and adults. Science, vol. 156(3777), pp. 986--988, (1967).

\bibitem{DSM}
Diagnostic and Statistical Manual of Mental Disorders, text revision (DSM-IV-TR), 4th Revision, American Psychiatric Association, (2000).

\bibitem{TF2006}
Fawcett, T., An introduction to ROC analysis. Pattern Recognition Letters, vol. 27, pp. 861--874, (2006).

\bibitem{GKK2013}
Gan, Y., Kakiashvili, T., Koczkodaj, W. W., Li, F., 
A note on relevance of diagnostic classification and rating scales used in psychiatry. Computer Methods and Programs Biomedicine, vol. 112(1), pp. 16--21, (2013).

\bibitem{Hair2006}
Hair, J. J., Black, W. C., Babin, B. J., Anderson, R. E., Tatham, R. L., Multivariate data analysis. New Jersey: Upper Saddle River, (2006).

\bibitem{ICD}
ICD-10: International Classification of Diseases. Geneva: World Health Organization, (1994).

\bibitem{KKW2011}
Kakiashvili, T., Koczkodaj, W. W., Woodbury-Smith, M., Improving the medical scale predictability by the pairwise comparisons method: evidence from a clinical data study,
Computer Methods and Programs Biomedicine, vol. 105(3), pp. 210--216, (2012).

\bibitem{KKL2014}
Koczkodaj, W. W., Kulakowski, K., Ligeza, A.,
On the quality evaluation of scientific entities in Poland supported by consistency-driven pairwise comparisons method. Scientometrics, vol. 99(3), pp. 911--926, (2014). 

\bibitem{Kocz96}  
Koczkodaj, W. W.,
Statistically Accurate Evidence of Improved Error Rate by Pairwise Comparisons. Perceptual and Motor Skills, vol. 82, pp. 43--48, (1996).

\bibitem{Kocz98}  
Koczkodaj, W.W., 
Testing the accuracy enhancement of pairwise comparisons by a Monte Carlo experiment,
Journal of Statistical Planning and Inference, vol. 69(1), pp. 21--31, 1998.

\bibitem{LBZO2005}
Lasko, T. A., Bhagwat, J. G., Zou, K. H., Ohno-Machado, L., The use of receiver operating characteristic curves in biomedical informatics, Journal of Biomedical Informatics, vol. 38(5), pp. 404-–415, (2005).

\bibitem{MR2012}
Le Moigne, P., Ragouet, P.,
Science as instrumentation. The case for psychiatric rating scales
Scientometrics, vol. 93(2), pp. 329-–349, (2012).

\bibitem{P2007}
Prpic, K.,
Changes of scientific knowledge production and research productivity in a transitional society Scientometrics, vol. 72(3), pp. 487--511, (2007).

\bibitem{R2006}
Rizopoulos, D.,
ltm: An R Package for Latent Variable Modelling and Item Response Theory Analyses. Journal of Statistical Software, vol. 17(5), (2006).

\bibitem{SSV2009}
Saris, W. E., Satorra, A., van der Veld, W. M., Testing Structural Equation Models or Detection of Misspecifications? Structural Equation Modeling: A Multidisciplinary Journal,
(2009).

\bibitem{sf}
$sourceforge.net/$ (accessed 2015-12-14).

\bibitem{WHO}
World Health Organization, (2013), Mental Health Action Plan 2013 - 2020, WHO Library Cataloguing-in-Publication Data
$http://www.who.int/mental_health/publications/action_plan/en/$ (accessed 2015-01-05).

\bibitem{ZOM2007}
Zou, K. H., O'Malley, A. J., Mauri, L., Receiver-operating characteristic analysis for evaluating diagnostic tests and predictive models. Circulation, vol. 115(5), pp. 654-–657, (2007).

\end{thebibliography}
\end{document}